\documentclass[a4paper,singlecolumn,twoside,12pt,showkeys]{revtex4}

\usepackage{amsmath,amssymb}	
\usepackage{bm}					
\usepackage{color}					
\usepackage{graphicx}				
\usepackage{dcolumn}				

\makeatletter
\def\frontmatter@abstractwidth{0.9\textwidth}	
\makeatother

\begin{document}
%
%
%


\newcommand{\By}{$\times$}
\newcommand{\SqrtBy}[2]{$\sqrt{#1}$\kern0.2ex$\times$\kern-0.2ex$\sqrt{#2}$}
\newcommand{\Degree}{$^\circ$}
\newcommand{\DegreeC}{$^\circ$C}
\newcommand{\Ohmcm}{$\Omega\cdot$cm}

\title{Effects of The Ehrlich-Schwoebel Potential Barrier
on the Wolf-Villain Model Simulations for Thin Film Growth}


\author{Rachan Rangdee}
\email{phongsaphat.ra@up.ac.th}
\author{Patcha Chatraphorn}
\affiliation{Department of Physics, Faculty of Science, Chulalongkorn University,
Bangkok, Thailand 10330}

\begin{abstract}

\vspace*{1mm}

\emph{Wolf-Villain (WV) model} is a simple model used to study \emph{ideal} molecular
beam epitaxy (MBE) growth by using computer simulations. In this model, an adatom
diffuses instantaneously within a finite diffusion length to maximize its coordination
number. We study statistical properties of thin films grown by this model. The morphology
of the WV model is found to be kinetically rough with a downhill particle diffusion current.
In real MBE growth, however, there are additional factors such as the existence of a
potential barrier that is known as the \emph{Ehrlich-Schwoebel (ES) barrier}. The ES
barrier is an additional barrier for an adatom that diffuses over a step edge from the
upper to a lower terrace which is known to induce an uphill particle current. We found
that with the addition of the ES barrier, the WV-ES model morphology is rough with
\emph{mound formation} on the surface when the barrier is strong enough. To confirm
these results, the correlation function is also studied. We find no oscillation in
the correlation function in the WV model. For the WV-ES model, the correlation function
oscillates. These results confirm that a strong enough ES barrier can cause mound formation
on the WV surface in our study.

\end{abstract}


\maketitle
\newpage


\section{Introduction}

Studies of the kinetic surface roughening models for molecular beam epitaxy (MBE)
growth has long been an interesting research topics
\cite{barabasi,wv,dt,ctd,ld,ljd,sw,ab,pd1,dpt,hhw,ya,ar}. This is because MBE technique
is very effective in growing high quality films. Scientists have tried to understand
behaviors of MBE growth from both theoretical and experimental points of view.
Many models have been proposed \cite{barabasi,wv,dt} as a simple
growth model for an \emph{ideal} MBE process which follows solid-on-solid (SOS)
constraints: desorption, overhanging and bulk vacancy are not allowed in the growing film.
In \emph{real} MBE growth, there are many factors effecting the film surfaces such as the
existence of a potential barrier known as the \emph{Ehrlich-Schwoebel (ES) barrier}
\cite{es1,es2}. The ES barrier is an additional potential barrier that prevents an
adatom to diffuse over a step edge from an upper to a lower terrace.

A substantial amount of works have been done to incorporate the ES barrier into MBE
growth model \cite{wv,dt,pd1,dpt,hhw,ya,ar}. One of the earliest works studied a stochastic
SOS Monte Carlo model in the presence of a potential barrier \cite{ld}. It was found that
the ``edge barrier'' induced mounding with a selected ``mound size'' because coarsening
become practically negligible after some time \cite{ld}. Since then there have been
many studies that use simple atomistic models to investigate effects of the ES barrier
on thin film growth \cite{sw,hhw,ya,dp}. All results confirm that the ES
barrier cause mound formation on the growing surfaces with the various mounds properties
depending on the details of the models and the ways that the barrier is incorporated
into the models.

In our work here, we use the Wolf-Villain (WV) model which was introduced in 1990 \cite{wv}
as a discrete limited mobility growth model for ideal low temperature growth under SOS
constraints. The ES barrier is added to see the effects of the barrier on the WV model.
Our work is done on a one dimensional substrate. This makes the situation more interesting
because the one dimensional WV model is known \cite{kps,dpt} to be in the Edward-Wilkinson
(EW) universality class \cite{ew} with a ``downhill particle current'' \cite{kps,dpt} that
stabilize the growth. The ES barrier, on the other hand, creates an ``uphill particle
current'' \cite{ctd,ya} that induces instability in the growing film. Therefore the goal
of our work is not just to try to understand the kinetic properties of the WV-ES model,
but we are also interested to see the interplay of the downhill current and uphill current
mechanisms when combined into one model.
\section{Theory}
To understand the kinetic surface roughening in MBE growth, we start by studying the
surface width which is defined as \cite{barabasi}
\begin{equation}
W(L,t) \equiv \langle(h - \bar{h})^2\rangle_x^{1/2},
\label{eqn1}
\end{equation}
where $h(x,t)$ is the surface height at any position $x$ and at time $t$ above a one
dimensional flat substrate of size $L$ lattice sites, $\bar{h}$ is the average height
of the surface and $\langle...\rangle_x$ is an average over the whole substrate. This
quantity let us know how rough our film surface is. The surface width depends on the
size of the substrate and growth time as \cite{barabasi,fv}
\begin{equation}
W(L,t) \sim \left\{ \begin{array}
                     {r@{\quad for \quad}l}
                     t^{\beta} & t \ll L^z \\ L^{\alpha} & t \gg L^z.
                    \end{array} \right.
\label{eqn2}
\end{equation}
Here $\beta$ is the \emph{growth} exponent, $\alpha$ is the \emph{roughness} exponent
and $z$ = $\alpha$/$\beta$ is the \emph{dynamical} exponent. These exponents ($\alpha$,
$\beta$,$z$) define the \emph{universality class} \cite{barabasi,dct} of the WV model.

There are two other powerful tools in studying the surface are the \emph{height-height
correlation function} \cite{ctd,dpt,dp,sp,fv} and the \emph{particle diffusion current}
\cite{kps}. First, we introduce the height-height correlation function which is defined as
\begin{equation}
G(r) = \langle h(x)h(x+r)\rangle_x,
\label{eqn3}
\end{equation}
where $h$ here represents the deviation of the surface height from an average height, i.e.
$h$ = $h$ - $\bar{h}$ and $r$ is the distance between two sites on the one dimensional
substrate. The correlation function can help us make a decision whether there is mound
formation on the surface. If the correlation function oscillates, it implies
\cite{ctd,dpt,dp,be} that there is a \emph{regular mound formation} on the surface. On
the other hand, if there is no oscillation in $G(r)$, there is no mound formation and the
surface is just kinetically rough. From the correlation function of the system with mounds,
we can also find the average mound height, $H$, from $H$ = $\sqrt{G(r = 0)}$. The average
mound radius, $R$, is the first zero crossing of the oscillating $G(r)$.

The particle diffusion current was introduced \cite{kps} as a tool for the study of a
conserved growth model. To measure the particle current $J$ in simulations, the growth
process is started on a tilted substrate. By ``tilted substrate'', it means the substrate
has a non-zero inclination $\tan\theta$ and the substrate height at any position $x$ and
initial time $t$ = 0 is set to be
\begin{equation}
h(x) = x \cdot \tan\theta,
\label{eqn8}
\end{equation}
for growth on one dimensional substrate (where the standard initial condition is $h(x)$ = 0).
We count the numbers of diffusing atoms during growth process
simulations. If an atom hops in the \emph{uphill(downhill)} direction, it contributes to
a positive(negative) current. For atoms that do not move (stick at the deposited sites)
after the deposition, the current from these atoms are zero. After the growth process is
completed, we calculate the net current by averaging overall adatoms deposited on the
surface. If the net current is positive the system is said to
have an \emph{uphill current} but if the net current is negative the system has a \emph{downhill
current}. For a system with an \emph{uphill} current, $J$ $>$ 0, we have an
unstable surface with mound formation or instabilities where as in a system with a
\emph{downhill} current, $J$ $<$ 0, the surface is stable and
belongs to the EW asymptotic universality class.

\section{Models and Methods}
\subsection{Wolf-Villain Model}
\begin{figure}[ht]
\begin{center}
\includegraphics[width=10cm]{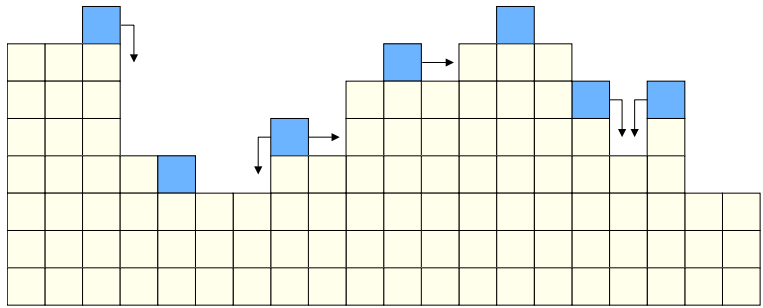}
\caption{Schematic diagram showing diffusion rules of the WV model on a one-dimensional
substrate.}
\end{center}
\label{fig1}
\end{figure}

The Wolf-Villain (WV) model \cite{wv} is a simple limited mobility model which was
introduced as a minimal model for the study of ideal MBE growth. By ``ideal'', it means
that there is no desorption, no bulk vacancy, and no overhanging. In other words, the
model follows solid-on-solid restrictions. In this model, an adatom is deposited on a
randomly chosen site on a one-dimensional flat substrate. Then the atom diffuses
instantaneously within a finite diffusion length $\ell$. In this diffusion process,
the adatom tries to diffuse to a site that offers the strongest bindings, i.e. the
atom tries to maximize its coordination number. Our study is based on $\ell$ = 1
situation which means the adatom deposited at site $x$ can only diffuse to its
nearest neighbor (NN) sites, $x$$\pm$1, as shown in Fig. \ref{fig1}.

In the simulation of the WV model, a deposition site $x$ is chosen randomly. We
then check the number of bondings at the deposition site ($n_x$) and compare it with
the number of bondings at NN sites ($n_{x\pm1}$). If $n_x$ $\ge$ $n_{x\pm1}$, which
means the NN sites do not have stronger bonding, then the atom is incorporated at
site $x$ and the height of site $x$ is increased by one. However, if one of the NN has
more bonding ($n_{x+1}$ $>$ $n_x$ or $n_{x-1}$ $>$ $n_x$), then the atom diffuses to that
NN site. Finally, if both NN have more bonding than at site $x$, then the atom goes to
the site with the strongest bonding or choose one of the NN sites with equal chance
if $n_{x+1}$ = $n_{x-1}$ $>$ $n_x$. The diffusion rule for the WV model is shown
schematically in Fig. \ref{fig1}. In all our simulations, periodic boundary condition
on a one-dimensional flat substrate is used.
\subsection{WV-ES Model}
\begin{figure}[ht]
\begin{center}
\includegraphics[width=10cm]{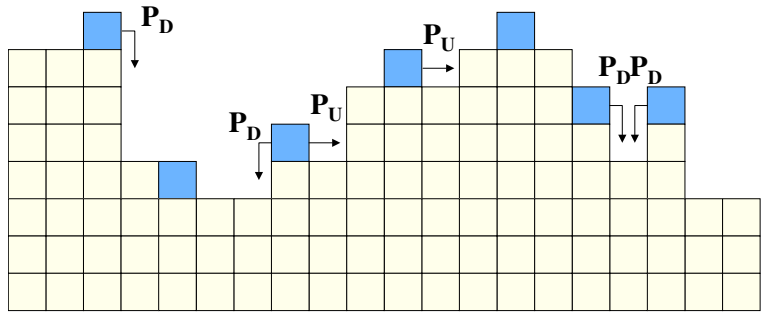}
\caption{Schematic diagram showing diffusion rules of the WV-ES model on a one-dimensional
substrate.}
\end{center}
\label{fig2}
\end{figure}
\begin{figure}[ht]
\begin{center}
\includegraphics[width=10cm]{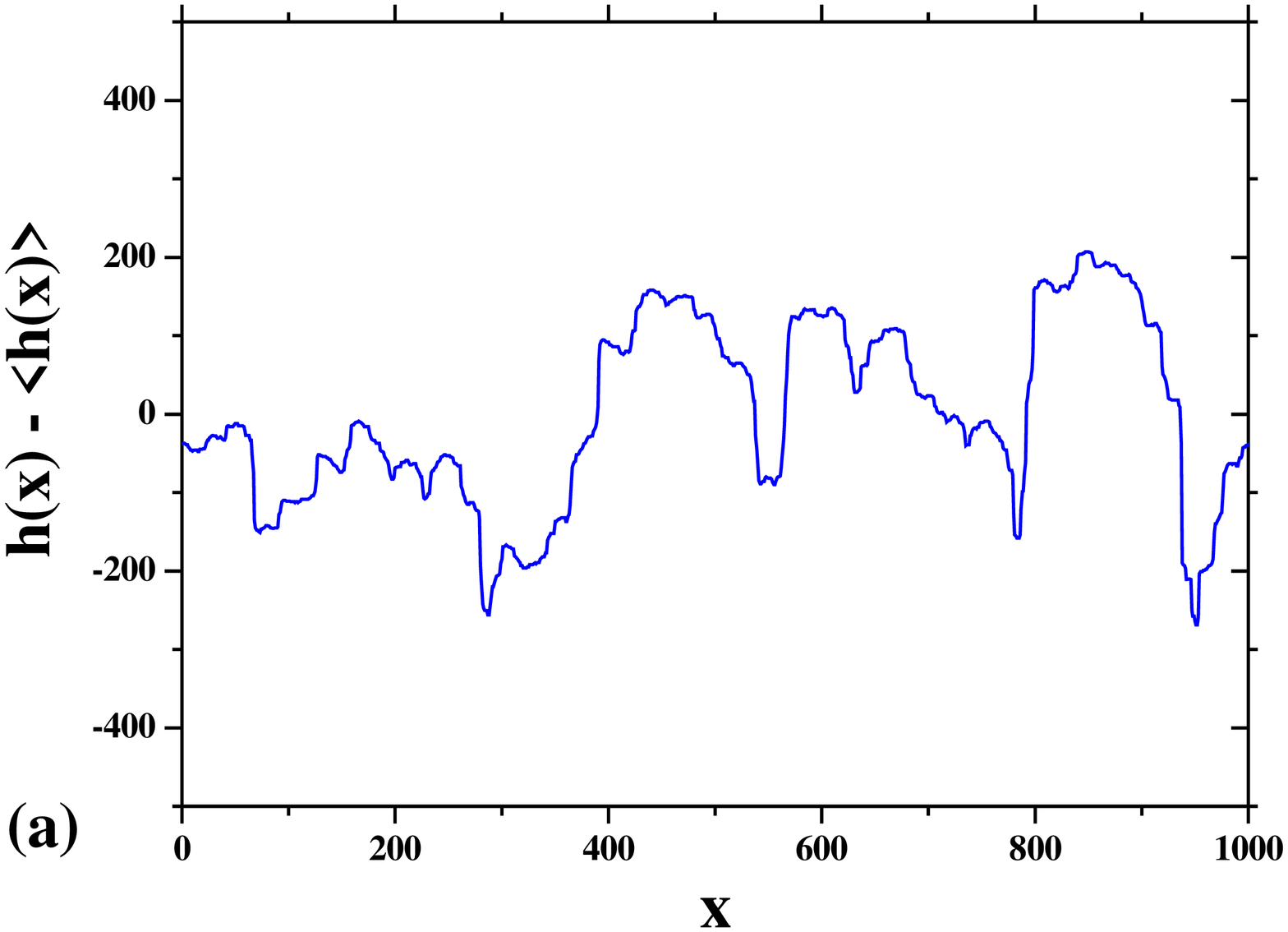}
\includegraphics[width=10cm]{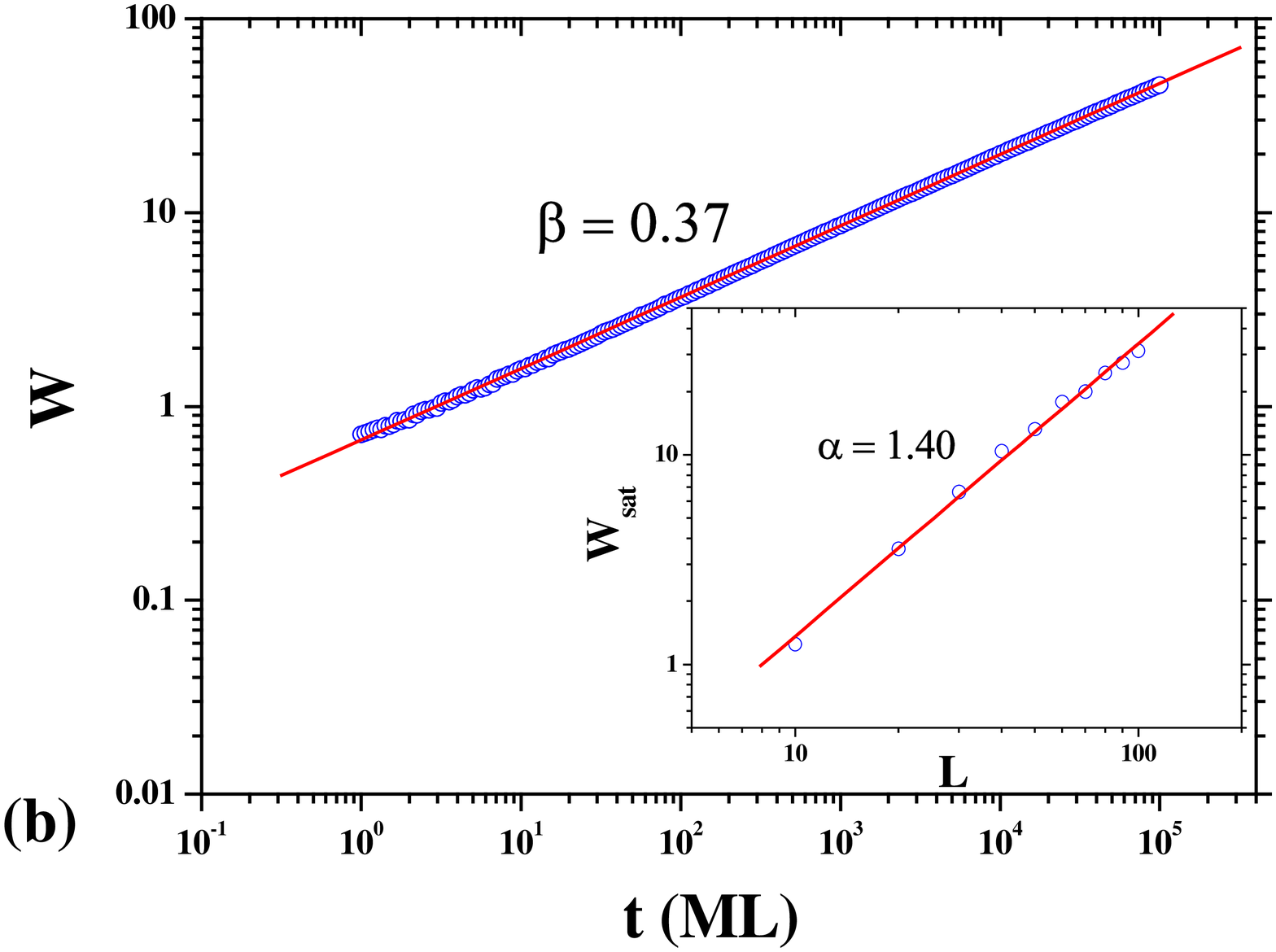}
\caption{(a) A typical WV morphology after depositing $10^6$ ML. (b) The surface width
versus time plot for the WV model ($L$ = 20,000). The inset shows the $W_{sat}$ as a
function of $L$.}
\end{center}
\label{fig3}
\end{figure}
From the original WV model \cite{wv}, we applied the effect of an ES barrier \cite{es1,es2}
by adding two probabilities $P_U$ and $P_D$. Here $P_U(P_D)$ is
a probability for an adatom to attach itself to the upper(lower) terrace after relaxation.
In Fig. \ref{fig2}, we show the modified WV model, the WV-ES model. The diffusion rule of
the WV-ES model is as follows: An atom is deposited on a randomly chosen site with the
average rate of 1 monolayer (ML) per second. The deposited atom looks for a better site
according to the original WV diffusion rule. The actual diffusion process is controlled
by the probabilities $P_U$ and $P_D$ in such a way that if the atom wants to diffuse to
the upper(lower) terrace, it faces the probability $P_U(P_D)$. If it cannot diffuse due
to the probability, it sticks at the original deposition site.

The ES barrier is implemented in our model by taking $P_U$ $>$ $P_D$ which makes it
more likely for an adatom to attach to the upper terrace than a lower terrace. The strength
of an ES barrier is controlled by the ratio $P_D$/$P_U$. If we set $P_U$ = $P_D$ = 1.0,
our model is back to the original WV model \cite{wv}.

\section{Results and Discussion}
Before showing our WV-ES results, we first briefly discuss the original WV model. Our WV
results are shown in Fig. \ref{fig3}. A typical morphology from our WV simulations after
$10^6$ ML deposition is shown in Fig. \ref{fig3}(a). It is clear from the morphology that
the interface of this model is kinetically rough. In Fig. \ref{fig3}(b), we show the surface
width $W$ plot as a function of time $t$ for a system with $L$ = 20,000 lattice sites. We
found the growth exponent to be $\beta$ $\approx$ 0.37, agreeing with previous works
\cite{ctd,wv,sk,ks,vv,hg,ky,rk}. Since $L$ is very large, we do not see any saturation
in this plot. The roughness exponent $\alpha$ is obtained from the plot of saturated surface
width ($W_{sat}$) as a function of substrate size $L$. In the inset of Fig. \ref{fig3}(b),
we plot the saturated width ($W_{sat}$) at time $t$ = $10^7$ ML versus the size of the
substrate $L$, with $L$ varying from $L$ = 10 to $L$ = 100. We obtained the value of
the roughness exponent in one-dimensional WV model to be $\alpha$ $\approx$ 1.40. Then
we calculated the third critical exponent (the dynamical exponent) $z$ by using the
relation $z$ = $\alpha$/$\beta$. The value of the dynamical exponent $z$ of the WV model
is $z$ $\approx$ 1.40/0.37 $\approx$ 4. All of the critical exponent ($\alpha$, $\beta$,
$z$) we obtained agree well with previous work \cite{ctd,wv,sk,ks,vv,hg,ky,rk}.
\begin{figure}[ht]
\begin{center}
\includegraphics[width=10cm]{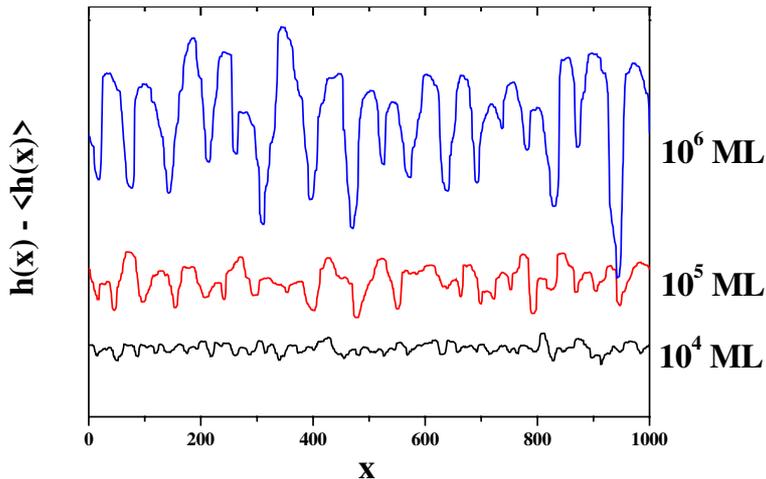}
\caption{The time evolution of the WV-ES morphologies. Here $L$ = 1000, $P_U$ = 1.0 and
$P_D$ = 0.5.}
\end{center}
\label{fig4}
\end{figure}

\begin{figure}[ht]
\begin{center}
\includegraphics[width=10cm]{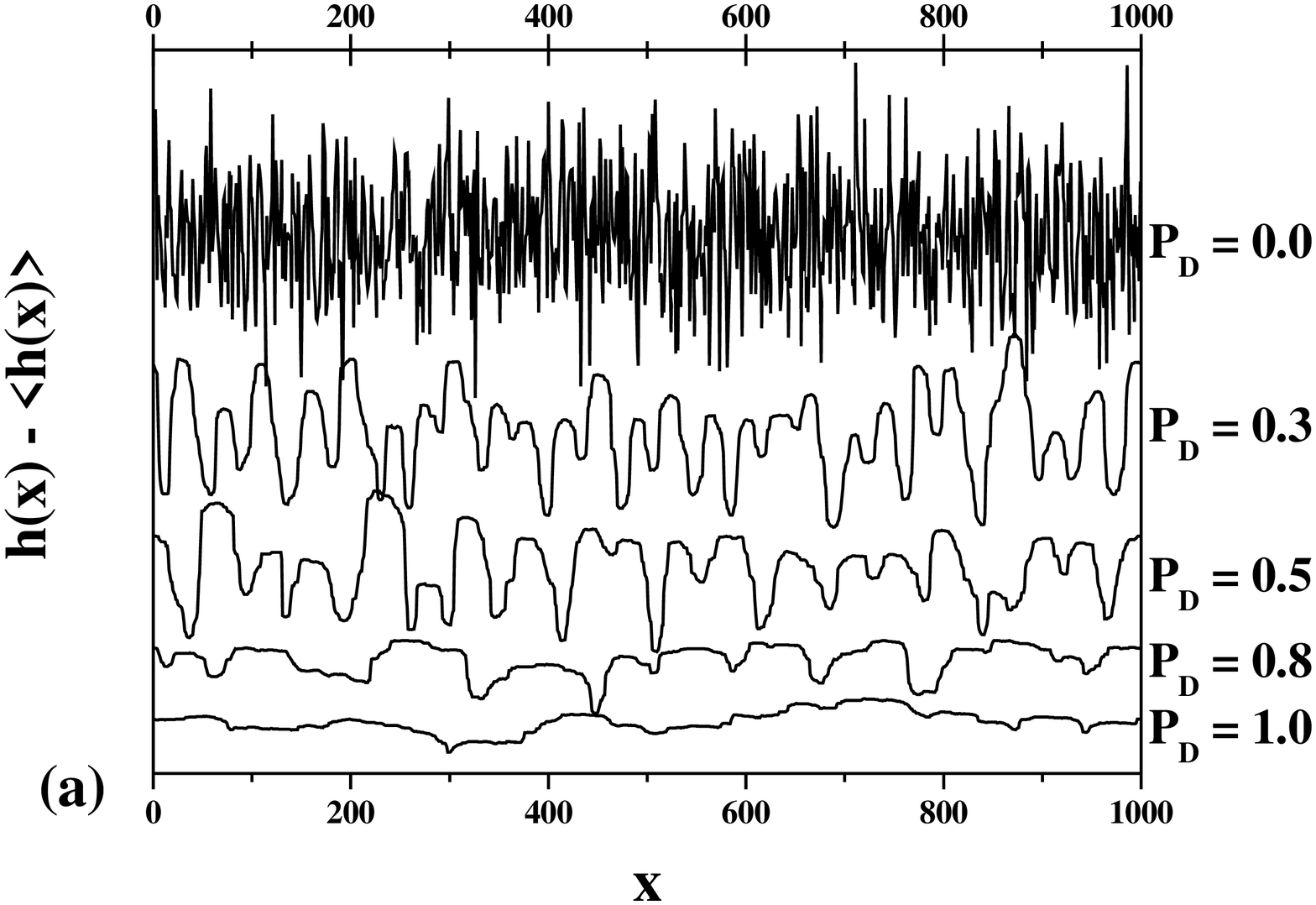}
\includegraphics[width=10cm]{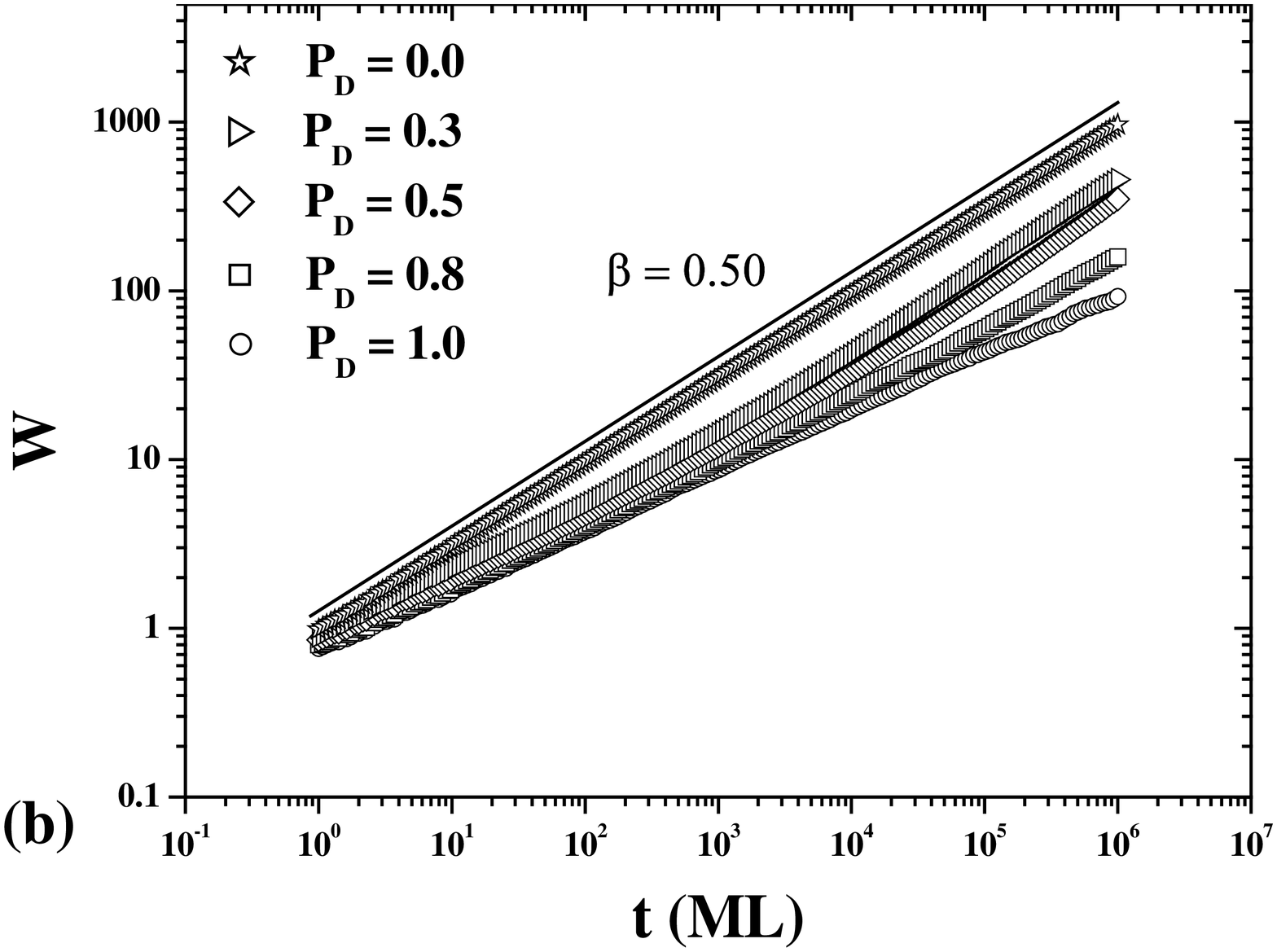}
\caption{(a) The morphologies of the WV-ES model at $t$ = $10^6$ ML. $P_U$ is fixed at 1.0
and $P_D$ is varied from 0.0 to 1.0. (b) The $W$-$t$ plot of the WV-ES model with $L$ =
1000, $P_U$ = 1.0.}
\end{center}
\label{fig5}
\end{figure}

Next we present our WV-ES results. The time evolution of the surface morphology of
the WV-ES model with $P_U$ = 1.0 and $P_D$ = 0.5 is shown in Fig. \ref{fig4}. Here, a
regular mounded pattern can clearly be seen at all three times. Mound coarsening (a process
where smaller mounds merge to become one larger mound) can also be seen, as there is more
mounds at $10^4$ ML compares with at $10^6$ ML. When we fixed $P_U$ = 1.0 and varied the
strength of the barrier by varying the probability $P_D$, we found that the surface
morphologies have deeper grooves and shaper peaks in systems with stronger barrier
(decreasing $P_D$), as in Fig. \ref{fig5}(a). This is due to the fact that adatoms have
less chance to hop down to the lower terraces when $P_D$ is smaller, which corresponds
to stronger barrier in this situation. The average size of each mound also seems to be
smaller when the barrier is stronger because coarsening is very slow. Note that when
$P_D$ = 1.0, the model is reduced back to the original WV model while $P_D$ = 0.0 is the
absolute barrier case where hopping down is not allowed at all. Corresponding $W$-$t$
plots for the WV-ES with varying barrier strength are shown in Fig. \ref{fig5}(b). The
bottom line, $P_U$ = $P_D$ = 1.0, is actually exactly the same as what was shown in Fig.
\ref{fig3}(a) as it is the same WV model \cite{wv}. When the ES barrier is added, weakly
at first, the $P_D$ = 0.8 system shows similar behavior as the WV model during early
time. However, as time increases, the surface width starts growing faster than in the WV
model. When the strength of the barrier is stronger, i.e. the value of $P_D$ is smaller,
it is obvious that after some time, the width increases at a faster rate and the growth
exponent of the WV-ES system is larger than that of the WV model. The value of $\beta$
in fact becomes much larger than 0.37 and it approaches the maximum possible value of
$\beta$ = 0.5. These results from the WV-ES model are remarkably similar to the results
from the SOS Monte Carlo model with edge barrier in the literature \cite{ld}. In both models,
regular mound formation is found and coarsening drastically slow down after some time
resulting in a large $\beta$ (Fig. \ref{fig5}(b)) and almost-constant correlation length
\cite{ld}.

For the absolute barrier $P_D$ = 0.0 situation, the growth exponent is
$\beta$ = 0.5 for the whole range of growth time. This is interesting because it means
the surface width of the WV model with absolute ES barrier case, which still allows
diffusion on the same layer, behaves in exactly the same way as the surface width of
the Random Deposition (RD) model \cite{barabasi} where no diffusion is allowed at all.
\begin{figure}[ht]
\begin{center}
\includegraphics[width=10cm]{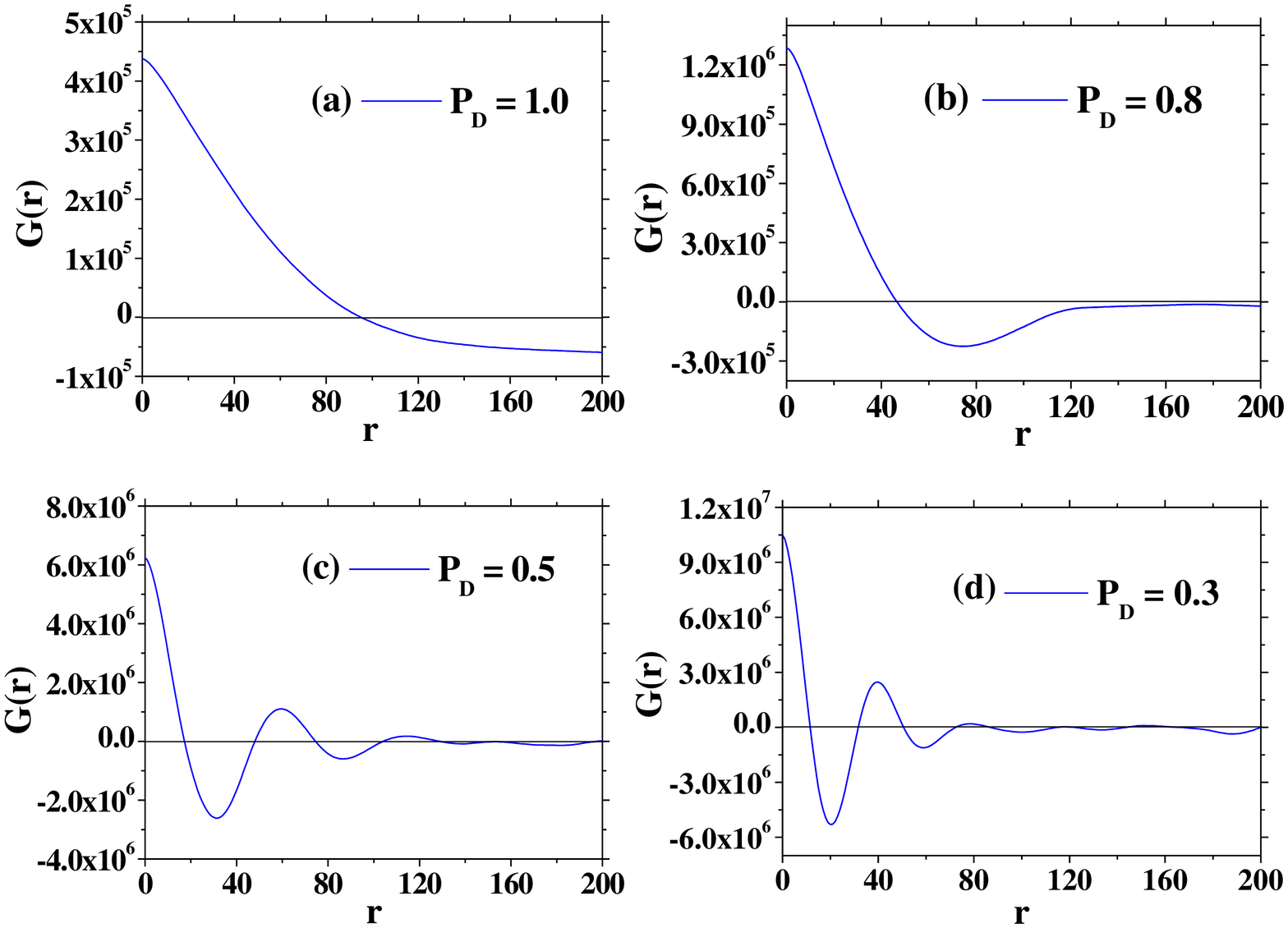}
\caption{The height-height correlation function of the WV-ES model with $L$ = 1000,
$t$ = $10^6$ ML and $P_U$ = 1.0. $P_D$ is varied.}
\end{center}
\label{fig6}
\end{figure}
\begin{figure}[p]
\begin{center}
\includegraphics[width=8.5cm]{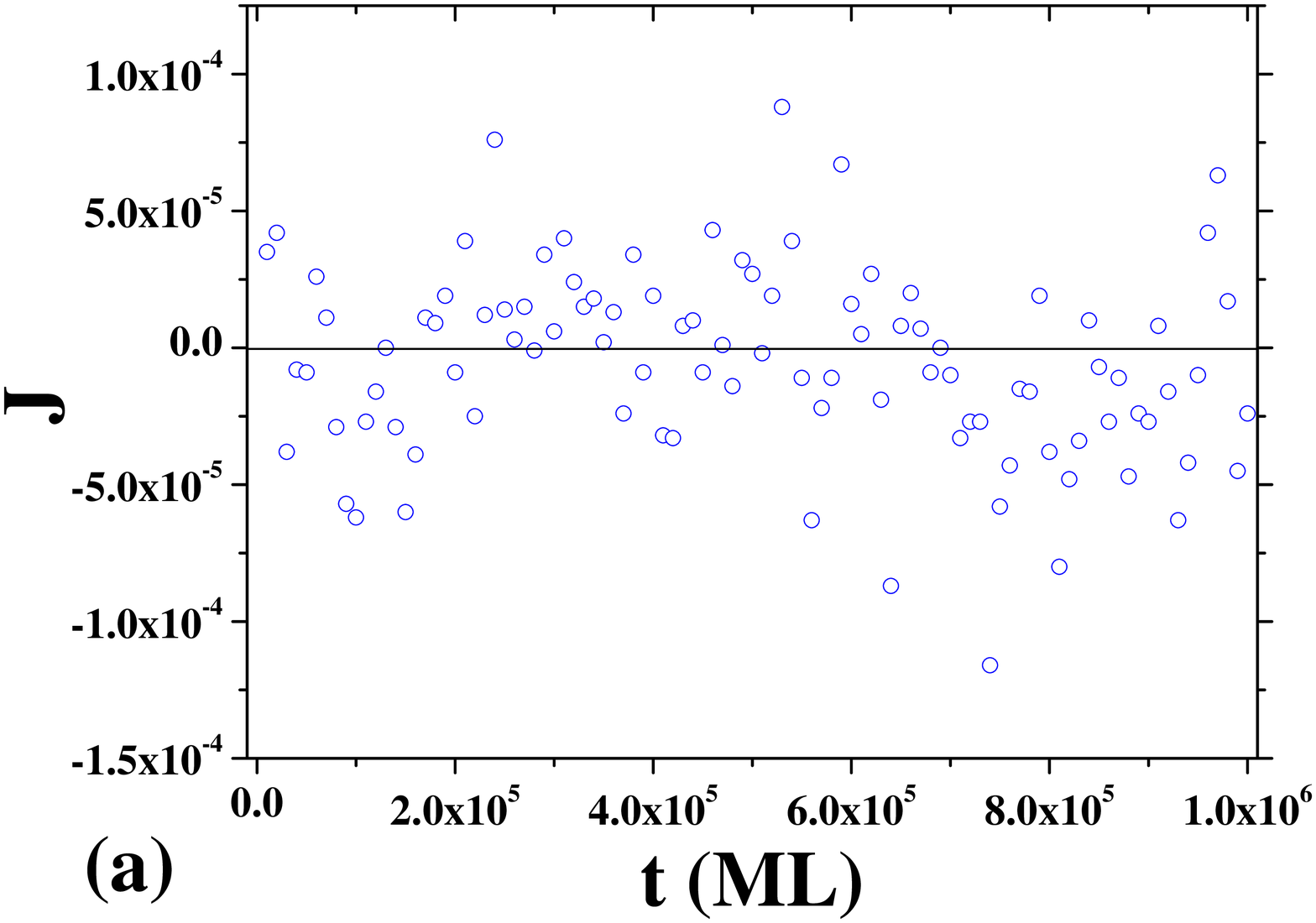}
\includegraphics[width=8.5cm]{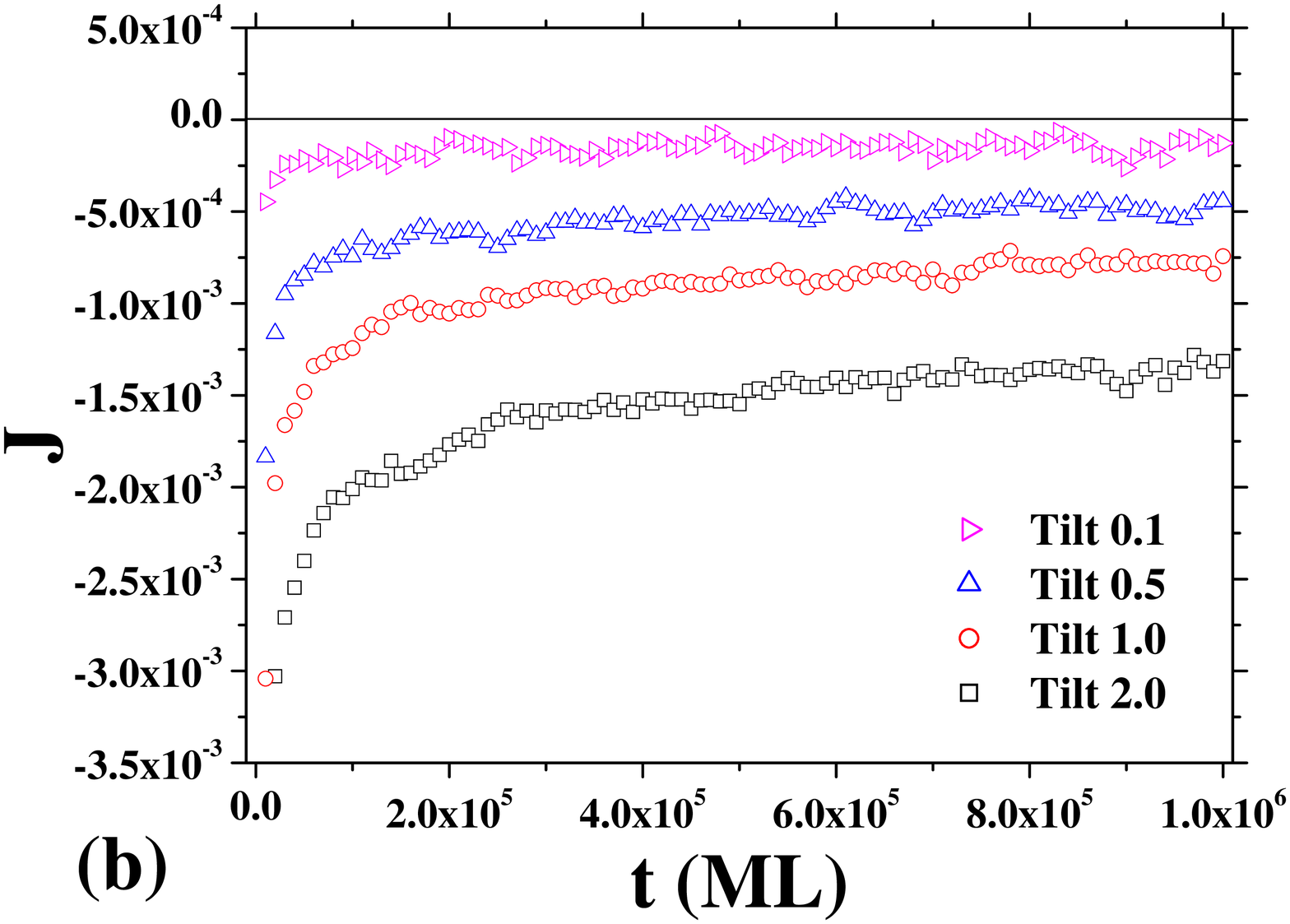}
\caption{The particle diffusion current of the WV model with untilted substrate in (a) and
vary tilted substrate in (b).}
\end{center}
\label{fig7}
\end{figure}
\begin{figure}[p]
\begin{center}
\includegraphics[width=10cm]{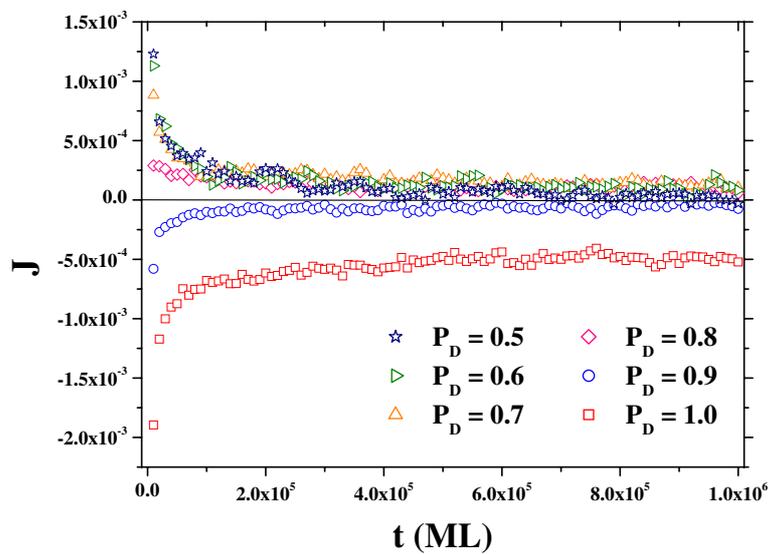}
\caption{The particle diffusion current of the WV-ES model by vary $P_D$ with
tan $\theta$ = 0.5.}
\end{center}
\label{fig8}
\end{figure}
\begin{figure}[p]
\begin{center}
\includegraphics[width=8.5cm]{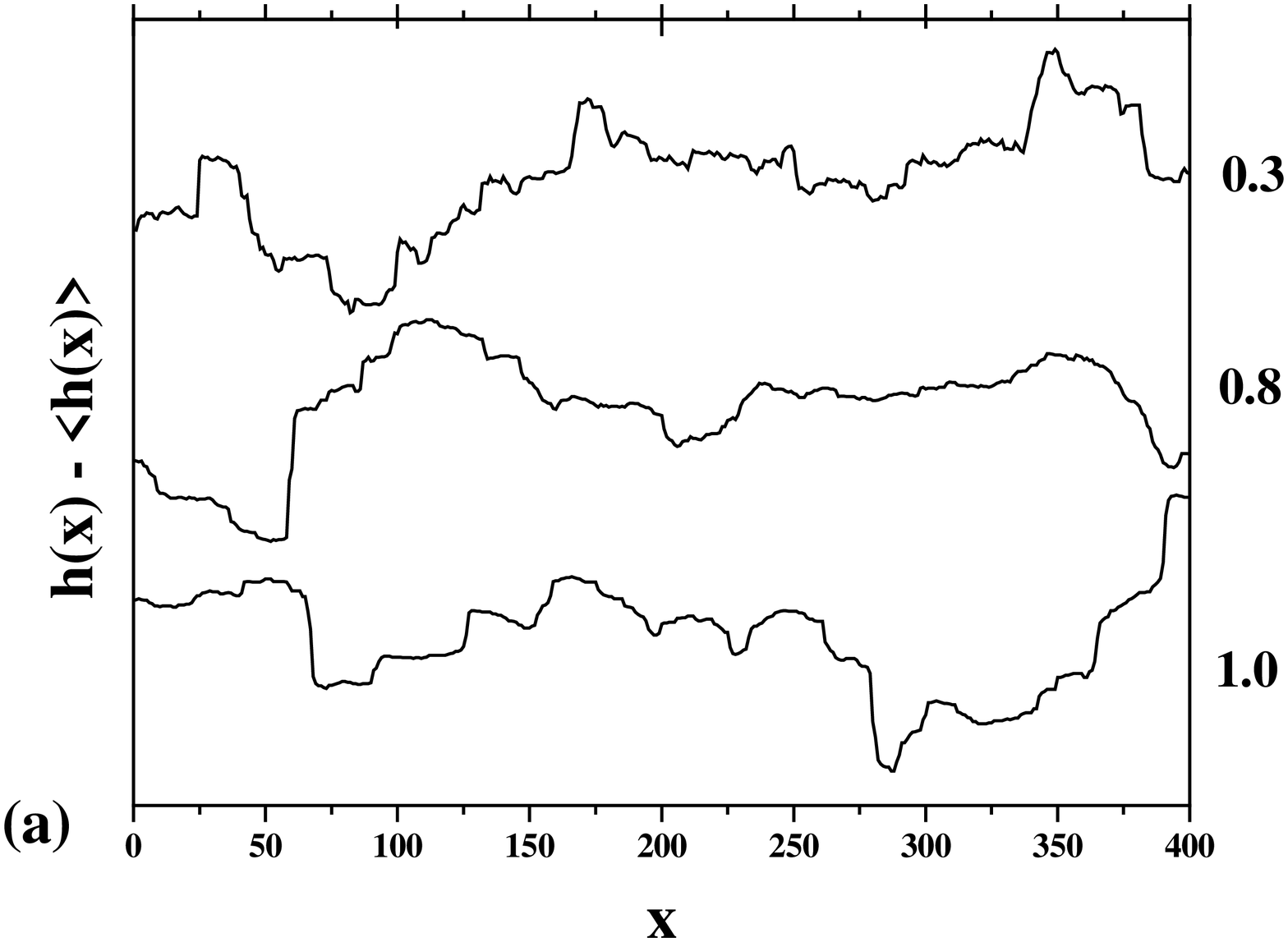}
\includegraphics[width=8.5cm]{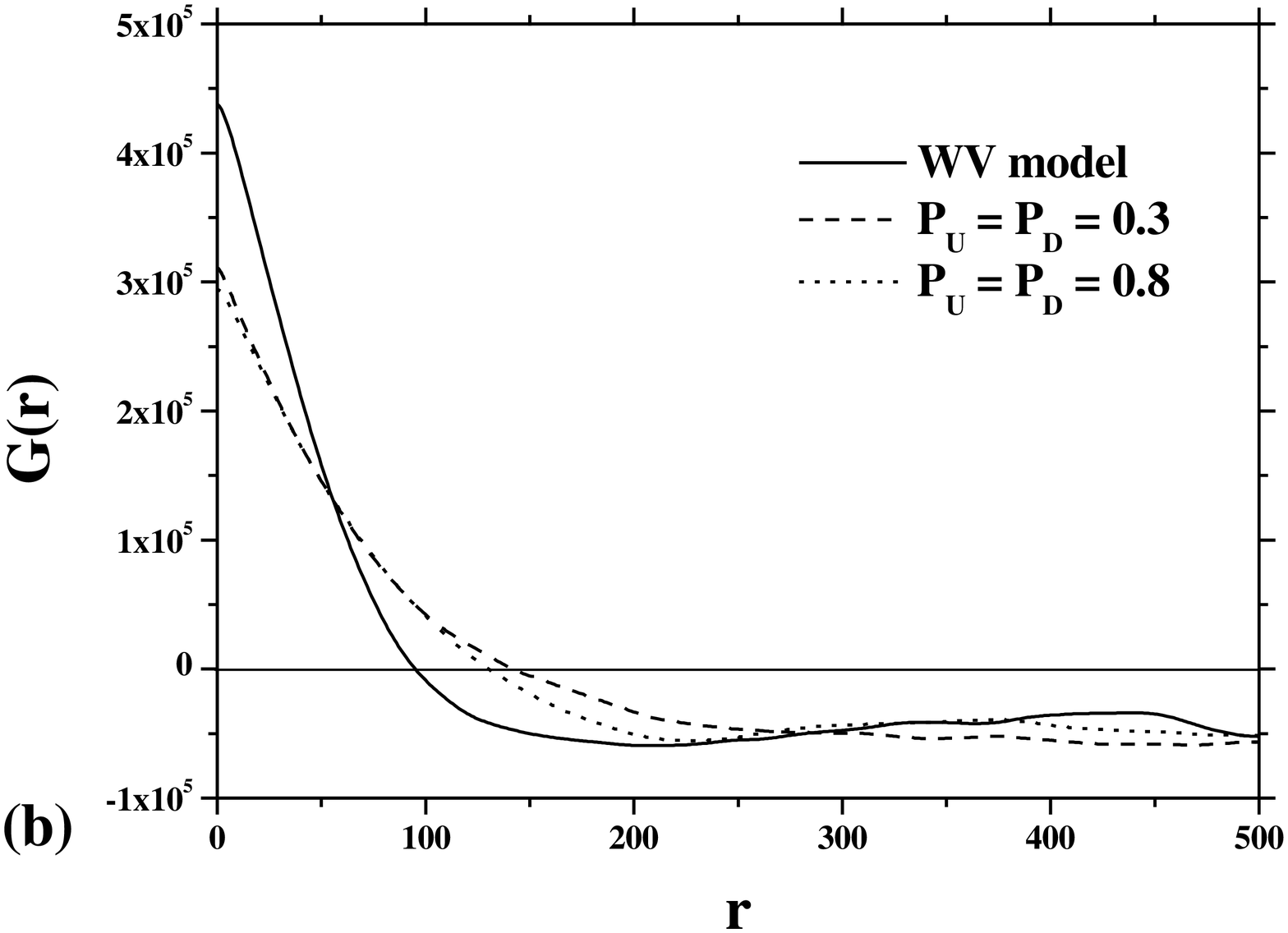}
\includegraphics[width=8.5cm]{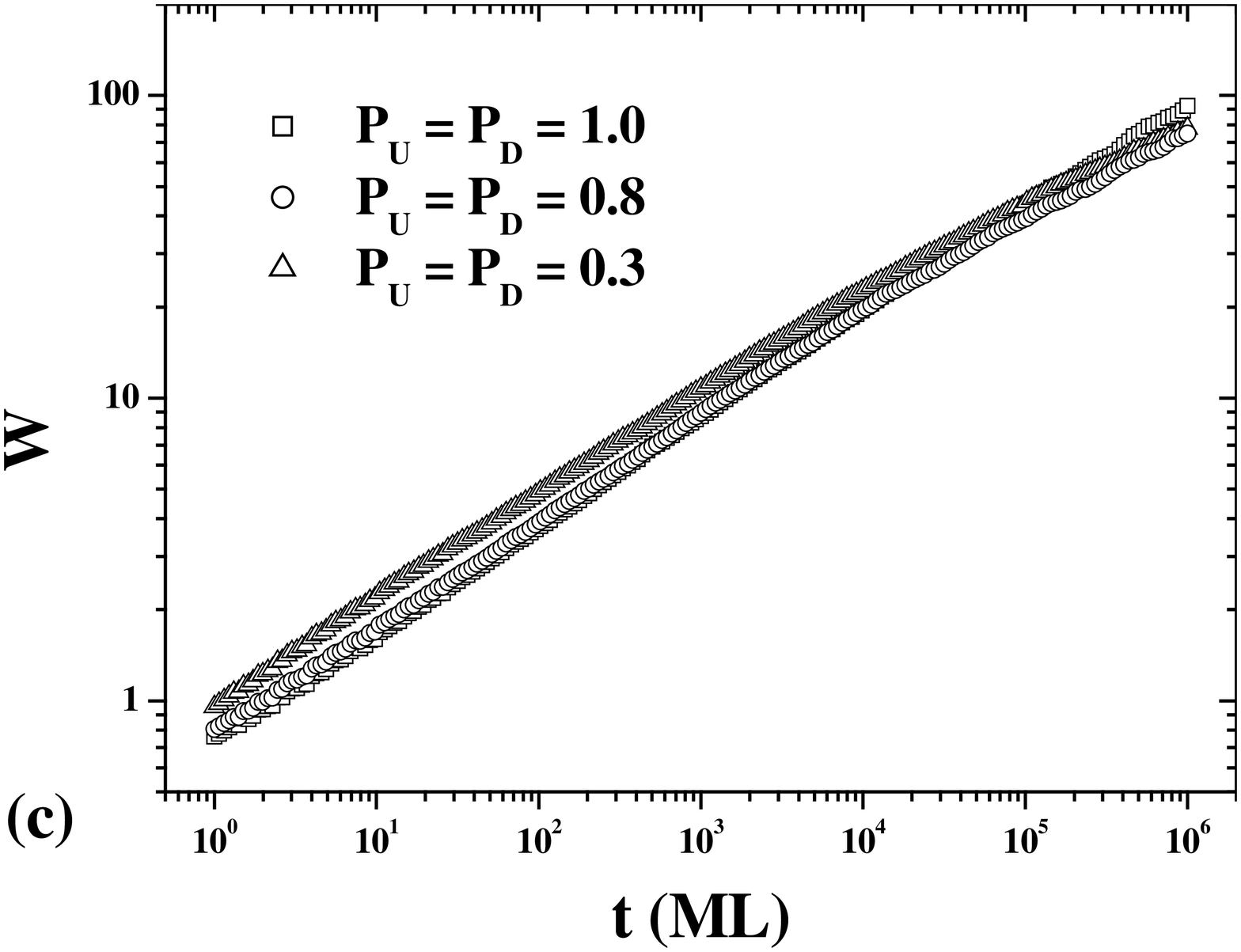}
\includegraphics[width=8.5cm]{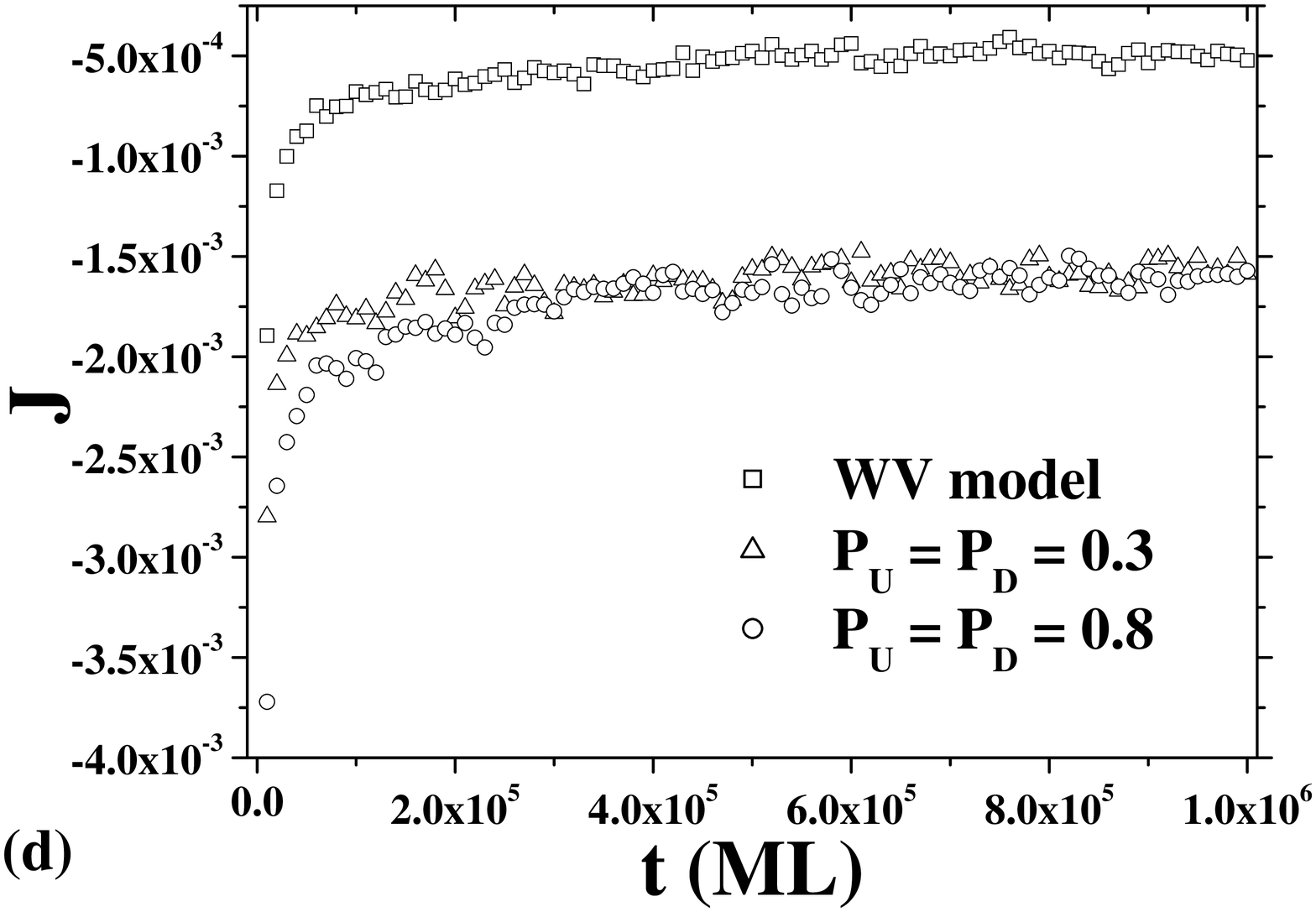}
\caption{Properties of the WV-ES model when $P_U$ = $P_D$. The morphologies are shown
in (a), the $G(r)$-$r$ plots are in (b), the $W$-$t$ plots are in (c), and the $J$-$t$
plots are in (d).}
\end{center}
\label{fig9}
\end{figure}

From our WV-ES results, it is evident that the ES barrier induces change in
statistical properties of the surface width of the WV model. The morphology also
changes from dynamically rough surface to regular mound formation. To investigate the
mounded surface further, we study the correlation function and the particle diffusion
current of the models. As described in the previous section, an oscillation in the
correlation function and uphill current are indicates of
regular mound pattern on the surface. Our $G(r)$ results are shown in Fig. \ref{fig6}.
In Fig. \ref{fig6}(a), the correlation function is calculated from the WV model without
the barrier. We have already showed that there is no mound in the WV morphology. Here, our
$G(r)$ results confirm it as there is no oscillation in the plot. Once we start adding
the ES barrier by fixing $P_U$ = 1.0 and decreasing $P_D$ by 0.1 at a time, we see that
the correlation functions start oscillating from $P_D$ = 0.7 (not shown) and smaller.
For $P_D$ = 0.8 (Fig. \ref{fig6}(b)) and $P_D$ = 0.9 (not shown), $G(r)$ behaves in the
same way as in the no-barrier model. When the barrier becomes stronger, the oscillation
is more pronounced as can be seen  when we set $P_D$ = 0.5 (Fig. \ref{fig6}(c)) and
$P_D$ = 0.3 (Fig. \ref{fig6}(d)). Note that the first zero crossing in Fig. \ref{fig6}(d)
is smaller than that in Fig. \ref{fig6}(c) which means the average mound radius in a
system with stronger barrier is smaller, agreeing with our morphology results in Fig.
\ref{fig5}(a).

Our results from the particle diffusion current are shown in Fig. \ref{fig7}. In Fig.
\ref{fig7}(a), the particle diffusion current is calculated from the original WV model
on a flat (untilt) substrate. So ``zero current'' in this case is defined to be in the
range between $\pm$5.0 $\times$ $10^{-5}$. From the tilt substrate systems, the net
current of the WV model is found to be \emph{negative} as
shown in Fig. \ref{fig7}(b). This result confirms that the WV model follows the EW
universality class asymptotically and it also implies that there is no mound formation
on the WV surface. Our downhill current in 1+1 WV model results agree with previous
works \cite{kps}.

For the WV-ES model, we fixed $P_U$ = 1.0 and then vary the values of $P_D$. We found that
for $P_D$ = 1.0 and 0.9, the net current is negative (downhill) while for $P_D$ $\le$
0.8 the net current is positive (uphill) as shown in Fig. \ref{fig8}. This indicates
that, at $P_D$ = 1.0 and 0.9, the barrier is too weak to effect the system so the
downhill current associating with the WV model has more influence on the system. The
surface in this case is dynamical rough without mound formation. However, when $P_D$
$\le$ 0.8 the ES barrier is strong and the uphill current associating with the ES
barrier has more influence. In this cases we obtain mound formation on the
surface (see Fig. \ref{fig5}(a) for the corresponding morphologies). Note that in the case
of $P_D$ = 0.8, there is a conflict between the correlation function study and the
particle diffusion current result, see Fig. \ref{fig6} and Fig. \ref{fig8}. This seems
to indicate that $P_D$ = 0.8 is the boundary between mound formation surface and dynamical
rough surface in the WV-ES model.

A comparison should be made here between the WV-ES results and the Das Sarma-Tamborenea
model with the ES barrier (DT-ES) results \cite{dp} as the WV model \cite{wv} and the
DT model \cite{dt} diffusion rules are very similar. Although our mounded results for
the WV-ES model with $P_D$ $<$ 0.8 are practically indistinguishable from the mounded
results for the 1+1 DT-ES model \cite{dp} in all aspects (e.g. morphologies, surface
width, coarsening rate, etc.), there is a major difference between the WV-ES and DT-ES
results. In the DT-ES model, there is no limit to the strength of the barrier. The
addition of the ES barrier, no matter how small, always creates mound formation \cite{dp}.
However, as we have shown here, the barrier has to be stronger than a certain value
to be able to induce mounds in the WV-ES model. The reason for this is that the WV \cite{wv}
and DT \cite{dt} models, though seemingly similar, belong to different universality classes.
The one dimensional WV model is in the EW universality with inherited downhill current
\cite{wv,kps,dct} so the ES barrier has to be strong enough to overcome the effect of
this negative current in the original WV model. On the other hand, the one dimensional
DT model has a zero particle current implying the absence of the EW term
\cite{kps,dct,pd2,slkg}. The zero current in the original DT model means that even a
small ES barrier still produces an uphill particle current and hence induces mounds
in the DT-ES model.

When studying effects of the ES barrier, we usually fix $P_U$ = 1.0 and vary the
value of $P_D$ in order to change the barrier strength. This can lead readers to think
that the absolute value of $P_U$ and $P_D$ are the controlling factor. This is, however,
not the case. the ES barrier is a \emph{``bias''} against the atom diffusion
in one direction (hopping down) when compared with diffusion in the other direction
(hopping on the same layer), so the absolute values of $P_U$ and $P_D$ are not as
important as the ratio $P_D$/$P_U$. To illustrate
this point, we simulated the WV-ES model with $P_U$ = $P_D$ $<$ 1.0. Our results are
shown in Fig. \ref{fig9}. We can see in Fig. \ref{fig9}(a) that the morphologies from
the systems with $P_U$ = $P_D$ = 0.3 and $P_U$ = $P_D$ = 0.8 are similar to the $P_U$ =
$P_D$ = 1.0 system which is just the regular WV model with no barrier. There is no
regular mound pattern on these surfaces even though the probabilities $P_U$ and $P_D$ are
less than unity. This is confirmed quantitatively by the correlation function results in
Fig. \ref{fig9}(b) where no oscillation is found in any of the systems. In Fig. \ref{fig9}(c),
we show the $W$-$t$ plots of the three systems and it is clear that they all have the same
statistical properties since the three lines intertwine together for the entire simulation
time which goes up to $10^6$ ML. Furthermore, we found downhill currents in all three
systems as shown in Fig. \ref{fig9}(d).

Finally, it should be noted that all our studies presented in this paper are results from
one dimensional substrate studies. The two dimensional WV-ES simulation was not done as the
results would certainly be mound formation for all values of the ES barrier strength. This
is because the  2+1 WV model, by itself, already exhibits mounded morphologies with selected
mound slope \cite{ctd,dct}. The mound formation in the two dimensional WV model is seen only
when measures to reduce background noise is taken \cite{ctd,dct} but the addition of the ES
barrier should enhance the uphill particle current already existed in the 2+1 WV model and
the mounds in 2+1 WV-ES model should be observed even without the noise reduction tecnique.

\section{Summary}
In our simulations, the WV model \cite{wv} shows dynamically rough surfaces with no mound
formation. This is confirmed by the calculation of the correlation function
\cite{ctd,dpt,dp,sp,fv} which does not oscillate. From the surface width, the critical
exponents are obtained to be $\beta$ $\approx$ 0.37, $\alpha$ $\approx$ 1.40, and $z$
$\approx$ 4. They agree well with other studies \cite{ctd,wv,sk,ks,vv,hg,ky,rk} in the same
time scale. In the simulations of the WV-ES model, we find regular mound formation on the
surface when the barrier is strong enough. In our study, the barrier has to be stronger than
$P_D$/$P_U$ = 0.8/1.0 in order to see mounds. The growth exponent increases from $\beta$
$\approx$ 0.37 in the WV model to approaches 0.5 in the mounded systems. We also find
oscillations in the correlation function and uphill current when the ES barrier is strong.
As the strength of the ES barrier increases, the average mound height increases and the
average mound radius decreases. We also find that when there is no bias in the direction
of the diffusion, i.e. $P_U$ = $P_D$, statistical behavior in these systems are the same
as in the original WV model \cite{wv} even when $P_U$ = $P_D$ $<$ 1.0.

\begin{acknowledgments}
The authors acknowledge financial supports from the grant for development of faculty
staff and the grant for graduate thesis from Chulalongkorn University, Bangkok, Thailand.
\end{acknowledgments}

\bibliographystyle{prb}
\bibliography{references}

\end{document}